\documentstyle[preprint,aps]{revtex}
\newtheorem{theorem}{Theorem}
\newtheorem{acknowledgement}[theorem]{Acknowledgement}

\begin{document}
\title{Effective inter-band coupling in $MgB_{2}$ due to anharmonic phonons.}
\author{M. Acquarone$^{1,2}$ and L. Roman\'{o}$^{2}$}
\address{$^{1}$IMEM-CNR, Parco Area delle Scienze, 43100 Parma (Italy); $^{2\text{ }}$%
Unit\`{a} INFM e Dipartimento di Fisica, Universit\`{a} di Parma, Parco Area%
\\
delle Scienze, 43100 Parma (Italy).}
\date{\today}
\maketitle

\begin{abstract}
We investigate the origin of the inter-band coupling in $MgB_{2}$ by
focusing on its unusual phononic features, namely, the strong anharmonicity
of the phonons and the presence of both linear and quadratic electron-phonon
interactions of the Su-Schrieffer-Heeger (SSH) type. The bare electronic
Hamiltonian has two bands with intra- and inter-band hopping, which lead to
two decoupled hybridized bands. The phonon Hamiltonian including the
anharmonic terms is diagonalized approximately by a squeezing
transformation, which causes the softening of the phonon frequency. The
linear SSH coupling amplitude is reduced, consistently with the estimates
from first-principle calculations. Additionally, the quadratic coupling
generates an effective phonon-induced interaction between the hybridized
bands, which is non-vanishing even in the limit of vanishing inter-bare-band
hopping amplitude. \\[1pt]

\end{abstract}

\section{Introduction.}

\bigskip The electronic structure of the $40$ K superconductor $MgB_{2}$ is
characterized by the presence at the Fermi level of two hybrid bands ($%
\sigma $ and $\pi )$ of very different character\cite{mazin}. This feature
reflects itself in the experimental evidence of two gaps, which, in the
absence of magnetic fields, have a common critical temperature\cite
{choi-nature}. This implies an interaction between the $\sigma $ and $\pi $
bands contributing to the Fermi surface\cite{suhl}. The microscopic origin
of this interaction is, to the best of our knowledge, not yet clarified. The
present work suggests that it might be traced to the unusual phononic
structure of $MgB_{2}.$ The dominant electron- phonon interaction is of the
Su-Schrieffer- Heeger (SSH) type, namely a modulation of the hopping due to
the vibration of the Boron ions along the inter-site bond\cite{dolgov}, \cite
{liu}. The unusual features are the presence of anharmonic contributions (up
to fourth order in the displacements) to the phononic Hamiltonian, and of
both \ a linear and a quadratic term in the SSH\ interaction\cite{liu}\cite
{Yildirim}\cite{choi}. Our aim is to give a qualitative picture of the issue
by discussing a physically transparent model, but with no ambition of giving
detailed quantitative results. However, we also show that our model is
quantitatively consistent with the numerical results of Refs.\cite{liu}\cite
{Yildirim}\cite{choi}

\section{The electronic Hamiltonian.}

\bigskip Our model of the electronic structure of $MgB_{2}$ by a Hamiltonian
has two bands, labelled $c$ and $d,$ which hybridize through an inter-site
hopping term. Then, in the real space representation, the bare electronic
Hamiltonian reads: 
\begin{equation}
H_{el}^{{}}=\varepsilon ^{c}\sum_{l\sigma }n_{l\sigma }^{c}+\varepsilon
^{d}\sum_{l\sigma }n_{l\sigma }^{d}+\sum_{l\langle j\rangle \sigma }\left[
t_{lj}^{cc}c_{l\sigma }^{\dagger }c_{j\sigma }^{{}}+t_{lj}^{dd}d_{l\sigma
}^{\dagger }d_{j\sigma }^{{}}\right] +\sum_{l\langle j\rangle \sigma
}t_{lj}^{cd}\left( c_{l\sigma }^{\dagger }d_{j\sigma }^{{}}+d_{j\sigma
}^{\dagger }c_{l\sigma }^{{}}\right)  \label{eq.Hel}
\end{equation}

where $\sum_{l\langle j\rangle }$ means summing on the $z$ nearest neigbors $%
j$ of a given site $l$, and then on $l.$ In $MgB_{2}$ one expects that $%
t_{lj}^{cc},t_{lj}^{dd}\gg t_{lj}^{cd}$ \cite{mazin},\cite{dolgov}. The
electron-phonon coupling parameters in the SSH scenario are derivatives of
the hopping amplitudes with respect to the inter-site distance. By using $%
c_{l\sigma }^{\dagger }=N^{-1/2}\sum_{k}c_{k\sigma }^{\dagger }\exp \left(
ikR_{l}^{{}}\right) $ we pass to the reciprocal space representation,
yielding : 
\begin{equation}
H_{el}^{{}}=\sum_{k\sigma }\left( \varepsilon ^{c}+zt_{k}^{cc}\right)
n_{k\sigma }^{c}+\sum_{k\sigma }\left( \varepsilon ^{d}+zt_{k}^{dd}\right)
n_{k\sigma }^{d}+\sum_{k\sigma }zt_{k}^{cd}\left( c_{k\sigma }^{\dagger
}d_{k\sigma }^{{}}+d_{k\sigma }^{\dagger }c_{k\sigma }^{{}}\right)
\end{equation}
where $t_{k}^{xy}=z^{-1}\sum_{\langle j\rangle }t_{lj}^{xy}\exp \left[
ik\left( R_{l}^{{}}-R_{j}^{{}}\right) \right] $ \ with $x,y=c,d$. To
diagonalize $H_{el}^{{}}$ we express the bare operators \ $c_{k\sigma
}^{\dagger },d_{k\sigma }^{\dagger }$\ through the hybridized operators $%
\alpha _{k\sigma }^{\dagger },\beta _{k\sigma }^{\dagger }$ according to: 
\begin{equation}
c_{k\sigma }^{\dagger }=\alpha _{k\sigma }^{\dagger }\cos \varphi
_{k}^{{}}+\beta _{k\sigma }^{\dagger }\sin \varphi _{k}^{{}}\qquad
d_{k\sigma }^{\dagger }=-\alpha _{k\sigma }^{\dagger }\sin \varphi
_{k}^{{}}+\beta _{k\sigma }^{\dagger }\cos \varphi _{k}^{{}}
\label{eq.hybrid}
\end{equation}

By choosing 
\begin{equation}
\tan \left( 2\varphi _{k}^{{}}\right) =-\frac{2zt_{k}^{cd}}{\varepsilon
_{k}^{c}-\varepsilon _{k}^{d}+z\left( t_{k}^{cc}-t_{k}^{dd}\right) }
\label{lin.7}
\end{equation}

$H_{el}^{{}}$ is brought to diagonal form $H_{el}^{{}}=\sum_{k\sigma }\left(
E_{k}^{\alpha }n_{k\sigma }^{\alpha }+E_{k}^{\beta }n_{k\sigma }^{\beta
}\right) $ , with the particle energies in the hybridized bands given by: 
\begin{equation}
E_{k}^{\alpha }=\frac{1}{2}\left[ \varepsilon _{{}}^{c}+\varepsilon
_{{}}^{d}+z\left( t_{k}^{cc}+t_{k}^{dd}\right) \right] +\frac{1}{2}\sqrt{%
\left[ \varepsilon _{{}}^{c}-\varepsilon _{{}}^{d}+z\left(
t_{k}^{cc}-t_{k}^{dd}\right) \right] ^{2}+\left( 2zt_{k}^{cd}\right) ^{2}}
\label{lin.8}
\end{equation}
\begin{equation}
E_{k}^{\beta }=\frac{1}{2}\left[ \varepsilon _{{}}^{c}+\varepsilon
_{{}}^{d}+z\left( t_{k}^{cc}+t_{k}^{dd}\right) \right] -\frac{1}{2}\sqrt{%
\left[ \varepsilon _{{}}^{c}-\varepsilon _{{}}^{d}+z\left(
t_{k}^{cc}-t_{k}^{dd}\right) \right] ^{2}+\left( 2zt_{k}^{cd}\right) ^{2}}
\label{eq.hybE}
\end{equation}

\section{ The phononic Hamiltonian with anharmonic terms.}

Following \cite{Yildirim}, \cite{choi} we assume that the purely phononic
Hamiltonian $H_{ph}^{{}}$ for $MgB_{2}$ has to include, apart from the usual
harmonic term, also a non-negligible quartic contribution. It takes the
form: 
\begin{equation}
H_{ph}^{{}}=\sum_{q}\frac{P_{q}^{{}}P_{-q}^{{}}}{2M}+\frac{M}{2}%
\sum_{q}\Omega _{q}^{2}u_{q}^{{}}u_{-q}^{{}}+\frac{M^{2}}{4}%
\sum_{qp}x_{qp}^{{}}\Omega _{q}^{2}\Omega
_{p}^{2}u_{q}^{{}}u_{-q}^{{}}u_{p}^{{}}u_{-p}^{{}}
\end{equation}

where $M$ \ is the Boron mass and $\Omega _{q}^{{}}$ is the frequency, at
the wavevector $q$ along the $\Gamma -A$ line, of the optical mode of $%
E_{2g} $ symmetry. The parameter $x_{qp}^{{}}$ expresses the strength of the
quartic term involving the wavevectors $\pm q$ and $\pm p.$ In $MgB_{2}$,
from Ref.\cite{Yildirim}, one can estimate $x_{qp}^{{}}\approx 7.8$ eV$%
^{-1}. $

By quantizing the phonon field according to the usual relations: 
\begin{equation}
u_{q}^{{}}=\sqrt{\frac{\hbar }{2M\Omega _{q}^{{}}}}\left( b_{-q}^{\dagger
}+b_{q}^{{}}\right) \qquad P_{q}^{{}}=i\sqrt{\frac{\hbar \Omega _{q}^{{}}}{2M%
}}\left( b_{-q}^{\dagger }-b_{q}^{{}}\right) \qquad L_{q}^{{}}=\sqrt{\frac{%
\hbar }{2M\Omega _{q}^{{}}}}  \label{ph.quant}
\end{equation}

the harmonic part becomes $\sum_{q}\hbar \Omega _{q}^{{}}\left(
b_{q}^{\dagger }b_{q}^{{}}+\frac{1}{2}\right) $. When quantizing the quartic
term, we neglect the terms with different numbers of creation and
destruction operators and keep the remaining four-operator products only
when diagonal. Namely, we approximate $b_{-q}^{\dagger }b_{q}^{\dagger
}b_{-p}^{{}}b_{p}^{{}}\approx \left( \delta _{p,q}^{{}}+\delta
_{p,-q}^{{}}\right) \nu _{q}^{{}}\nu _{-q}^{{}}$ , where $b_{q}^{\dagger
}b_{q}^{{}}=\nu _{q}^{{}}$. The quartic contribution then reduces to: 
\[
\sum_{qp}x_{qp}^{{}}\left( \frac{\hbar \Omega _{q}^{{}}}{4}\right) \left( 
\frac{\hbar \Omega _{p}^{{}}}{4}\right) \left( b_{-q}^{\dagger
}+b_{q}^{{}}\right) \left( b_{q}^{\dagger }+b_{-q}^{{}}\right) \left(
b_{-p}^{\dagger }+b_{p}^{{}}\right) \left( b_{p}^{\dagger
}+b_{-p}^{{}}\right) \approx 
\]

\[
\approx 4\sum_{q}\left( \frac{\hbar \Omega _{q}^{{}}}{4}\right) \left( \frac{%
1}{2}+\nu _{q}^{{}}\right) \sum_{p}x_{qp}^{{}}\left( \frac{\hbar \Omega
_{p}^{{}}}{4}\right) \left( 1+\delta _{qp}^{{}}\right) 
\]
\[
+4\sum_{qp}x_{qp}^{{}}\left( \frac{\hbar \Omega _{p}^{{}}}{4}\right) \left( 
\frac{\hbar \Omega _{q}^{{}}}{4}\right) \nu _{q}^{{}}\nu _{p}^{{}}\left(
1+\delta _{q,-p}^{{}}\right) +2\sum_{q}\left( \frac{\hbar \Omega _{q}^{{}}}{4%
}\right) \left( b_{-q}^{\dagger }b_{q}^{\dagger
}+b_{-q}^{{}}b_{q}^{{}}\right) \sum_{p}x_{qp}^{{}}\left( \frac{\hbar \Omega
_{p}^{{}}}{4}\right) 
\]
\begin{equation}
-2\sum_{qp}x_{qp}^{{}}\left( \frac{\hbar \Omega _{q}^{{}}}{4}\right) \left( 
\frac{\hbar \Omega _{p}^{{}}}{4}\right) +\sum_{qp}x_{qp}^{{}}\left( \frac{%
\hbar \Omega _{q}^{{}}}{4}\right) \left( \frac{\hbar \Omega _{p}^{{}}}{4}%
\right)  \label{5.2}
\end{equation}

The product $\nu _{q}^{{}}\nu _{p}^{{}}$\ is approximated in the MFA
fashion, i.e. $\nu _{q}^{{}}\nu _{p}^{{}}\approx \nu _{q}^{{}}\langle \nu
_{p}^{{}}\rangle +\langle \nu _{q}^{{}}\rangle \nu _{p}^{{}}-\langle \nu
_{p}^{{}}\rangle \langle \nu _{q}^{{}}\rangle $ . Putting together the
constant terms, we can rewrite Eq.\ref{5.2} as: 
\[
\sum_{q}\hbar \Omega _{q}^{{}}\left( \frac{1}{2}+\nu _{q}^{{}}\right)
\sum_{p}x_{qp}^{{}}\left( \frac{\hbar \Omega _{p}^{{}}}{2}\right) \left( 
\frac{1}{2}+\langle \nu _{p}^{{}}\rangle \right) \left( 1+\delta
_{q,-p}^{{}}\right) 
\]
\begin{equation}
+\sum_{q}\hbar \Omega _{q}^{{}}\left( b_{-q}^{\dagger }b_{q}^{\dagger
}+b_{-q}^{{}}b_{q}^{{}}\right) \sum_{p}x_{qp}^{{}}\left( \frac{\hbar \Omega
_{p}^{{}}}{8}\right) +{\rm const.}  \label{8.1}
\end{equation}

Adding the harmonic contribution and defining 
\begin{equation}
X_{q}^{{}}\equiv 1+\sum_{p}x_{qp}^{{}}\left( \frac{\hbar \Omega _{p}^{{}}}{2}%
\right) \left( \frac{1}{2}+\langle \nu _{p}^{{}}\rangle \right) \left(
1+\delta _{q,-p}^{{}}\right)  \label{eqXq}
\end{equation}

we obtain the purely phononic Hamiltonian as: 
\begin{equation}
H_{ph}=\sum_{q}\hbar \Omega _{q}^{{}}X_{q}^{{}}\left( \frac{1}{2}+\nu
_{q}^{{}}\right) +\sum_{q}\hbar \Omega _{q}^{{}}\left( b_{-q}^{\dagger
}b_{q}^{\dagger }+b_{-q}^{{}}b_{q}^{{}}\right) \sum_{p}x_{qp}^{{}}\left( 
\frac{\hbar \Omega _{p}^{{}}}{8}\right) +{\rm const.}  \label{8.2b}
\end{equation}

This form can be diagonalized by a \ ''squeezing '' transformation\cite
{squeeze} $e^{S}\equiv \exp \left[ -\sum_{q}\eta _{q}^{{}}\left(
b_{-q}^{\dagger }b_{q}^{\dagger }-b_{-q}^{{}}b_{q}^{{}}\right) \right] $
under the condition that 
\begin{equation}
\tanh \left( 2\eta _{q}^{{}}\right) =-\frac{1}{X_{q}^{{}}}%
\sum_{p}x_{qp}^{{}}\left( \frac{\hbar \Omega _{p}^{{}}}{4}\right)
\label{10.4}
\end{equation}

Notice that Eq.\ref{10.4} yields $\eta _{q}^{{}}<0.$ The diagonalized
Hamiltonian $e^{S}H_{ph}e^{-S}$ can now be written as: 
\begin{equation}
e^{S}H_{ph}e^{-S}=\sum_{q}\hbar \Omega _{q}^{{}}\left[ X_{q}^{{}}\cosh
\left( 2\eta _{q}^{{}}\right) +2\sinh \left( 2\eta _{q}^{{}}\right)
\sum_{p}x_{qp}^{{}}\frac{\hbar \Omega _{p}^{{}}}{8}\right] \left(
b_{q}^{\dagger }b_{q}^{{}}+\frac{1}{2}\right) +{\rm const.}  \label{11.3}
\end{equation}

By substituting $\eta _{q}^{{}}$ from Eq.\ref{10.4} into Eq.\ref{11.3}, the
renormalized frequency $\omega _{q}^{{}}$ of the harmonic Hamiltonian for
the squeezed phonons is written explicitly as: 
\begin{equation}
\omega _{q}^{{}}=\Omega _{q}^{{}}X_{q}^{{}}\left[ \sqrt{1-\tanh ^{2}\left(
2\eta _{q}^{{}}\right) }\right]  \label{SQomega}
\end{equation}

where $\Omega _{q}^{{}}X_{q}^{{}}$ is the phonon frequency entering the
quadratic part of the unsqueezed phononic Hamiltonian (see. Eq.\ref{8.2b}).
Thus $\omega _{q}^{{}}$ is reduced (softened) with respect to the bare
frequency $X_{q}^{{}}\Omega _{q}^{{}}$, in accordance with the findings of
Refs.\cite{liu}\cite{Yildirim}\cite{choi}.

\section{The linear electron-phonon interaction.}

The linear part of the SSH\ electron-phonon interaction is written, in real
space, as 
\begin{equation}
H_{ep}^{(1)}=\sum_{l\langle j\rangle \sigma }\left[ g_{lj}^{cc}c_{l\sigma
}^{\dagger }c_{j\sigma }^{{}}+g_{lj}^{dd}d_{l\sigma }^{\dagger }d_{j\sigma
}^{{}}\right] \left( u_{l}^{{}}-u_{j}^{{}}\right) +\sum_{l\langle j\rangle
\sigma }g_{lj}^{cd}\left( d_{l\sigma }^{\dagger }c_{j\sigma
}^{{}}+c_{j\sigma }^{\dagger }d_{l\sigma }^{{}}\right) \left(
u_{l}^{{}}-u_{j}^{{}}\right)  \label{lin.1}
\end{equation}

where $g_{lj}^{cc}=\partial t_{lj}^{cc}/\partial \left(
u_{l}^{{}}-u_{j}^{{}}\right) |_{0}=-g_{jl}^{cc}$ etc., are the coupling
constants.

To obtain the Fourier-transformed form of Eq.\ref{lin.1} we define $%
g_{k}^{cc}=(1/z)\sum_{\langle j\rangle }g_{lj}^{cc}\exp (ik\Delta
_{lj}^{{}}) $ so that 
\begin{equation}
\frac{z}{2}\left( g_{k-q}^{cc}+g_{k}^{cc}-g_{-(k-q)}^{cc}-g_{-k}^{cc}\right)
=i\sum_{\langle j\rangle }g_{lj}^{cc}\left\{ \sin \left[ \left( k-q\right)
\cdot \Delta _{lj}^{{}}\right] -\sin \left[ k\cdot \Delta _{lj}^{{}}\right]
\right\} \equiv \gamma _{k,q}^{cc}  \label{lin.3}
\end{equation}

with analogous relations defining $\gamma _{k,q}^{dd}$ and $\gamma
_{k,q}^{cd}$. Quantization of the phonons according to Eq.\ref{ph.quant}
leads to: 
\begin{equation}
H_{ep}^{(1)}=\frac{1}{\sqrt{N}}\sum_{kq\sigma }L_{q}\left[ \gamma
_{k,q}^{cc}c_{k\sigma }^{\dagger }c_{k-q\sigma }^{{}}+\gamma
_{k,q}^{dd}d_{k\sigma }^{\dagger }d_{k-q\sigma }^{{}}+\gamma
_{k,q}^{cd}\left( c_{k\sigma }^{\dagger }d_{k-q\sigma }^{{}}+d_{k-q\sigma
}^{\dagger }c_{k\sigma }^{{}}\right) \right] \left( b_{-q}^{\dagger
}+b_{q}^{{}}\right)  \label{lin.4}
\end{equation}

When transformed to the hybridized fermion representation $H_{ep}^{(1)}$
reads: 
\[
H_{ep}^{(1)}= 
\]
\begin{equation}
=\frac{1}{\sqrt{N}}\sum_{kq\sigma }L_{q}\left[ \Gamma _{k,k-q}^{\alpha
\alpha }\alpha _{k\sigma }^{\dagger }\alpha _{k-q,\sigma }^{{}}+\Gamma
_{k,k-q}^{\beta \beta }\beta _{k\sigma }^{\dagger }\beta _{k-q,\sigma
}^{{}}+\Gamma _{k,k-q}^{\alpha \beta }\alpha _{k\sigma }^{\dagger }\beta
_{k-q,\sigma }^{{}}+\Gamma _{k,k-q}^{\beta \alpha }\beta _{k\sigma
}^{\dagger }\alpha _{k-q,\sigma }^{{}}\right] \left( b_{-q}^{\dagger
}+b_{q}^{{}}\right)  \label{lin.9}
\end{equation}

where the effective couplings are defined as: 
\begin{equation}
\Gamma _{k,k-q}^{\alpha \alpha }=\gamma _{k,k-q}^{cc}\cos \varphi
_{k}^{{}}\cos \varphi _{k-q}^{{}}+\gamma _{k,k-q}^{dd}\sin \varphi
_{k}^{{}}\sin \varphi _{k-q}^{{}}-2\gamma _{k,k-q}^{cd}\cos \varphi
_{k}^{{}}\sin \varphi _{k-q}^{{}}  \label{lin.10}
\end{equation}

\begin{equation}
\Gamma _{k,k-q}^{\beta \beta }=\gamma _{k,k-q}^{cc}\sin \varphi
_{k}^{{}}\sin \varphi _{k-q}^{{}}+\gamma _{k,k-q}^{dd}\cos \varphi
_{k}^{{}}\cos \varphi _{k-q}^{{}}+2\gamma _{k,k-q}^{cd}\sin \varphi
_{k}^{{}}\cos \varphi _{k-q}^{{}}  \label{lin.11}
\end{equation}

\begin{equation}
\Gamma _{k,k-q}^{\alpha \beta }=\gamma _{k,k-q}^{cc}\cos \varphi
_{k}^{{}}\sin \varphi _{k-q}^{{}}-\gamma _{k,k-q}^{dd}\sin \varphi
_{k}^{{}}\cos \varphi _{k-q}^{{}}+\gamma _{k,k-q}^{cd}\left( \cos \varphi
_{k}^{{}}\cos \varphi _{k-q}^{{}}+\sin \varphi _{k}^{{}}\sin \varphi
_{k-q}^{{}}\right)  \label{lin.12}
\end{equation}

\begin{equation}
\Gamma _{k,k-q}^{\beta \alpha }=\gamma _{k,k-q}^{cc}\sin \varphi
_{k}^{{}}\cos \varphi _{k-q}^{{}}-\gamma _{k,k-q}^{dd}\cos \varphi
_{k}^{{}}\sin \varphi _{k-q}^{{}}-\gamma _{k,k-q}^{cd}\left( \cos \varphi
_{k}^{{}}\cos \varphi _{k-q}^{{}}+\sin \varphi _{k}^{{}}\sin \varphi
_{k-q}^{{}}\right)  \label{lin.13}
\end{equation}

\section{The quadratic electron-phonon interaction.}

According to Refs.\cite{liu}\cite{Yildirim}\cite{choi}, the electron-phonon
Hamiltonian has to include also a quadratic term, which we write in real
space as: 
\begin{equation}
H_{ep}^{(2)}=\sum_{l\langle j\rangle \sigma }\left[ f_{lj}^{cc}c_{l\sigma
}^{\dagger }c_{j\sigma }^{{}}+f_{lj}^{dd}d_{l\sigma }^{\dagger }d_{j\sigma
}^{{}}\right] \left( u_{l}^{{}}-u_{j}^{{}}\right) ^{2}+\sum_{l\langle
j\rangle \sigma }f_{lj}^{cd}\left( d_{l\sigma }^{\dagger }c_{j\sigma
}^{{}}+c_{j\sigma }^{\dagger }d_{l\sigma }^{{}}\right) \left(
u_{l}^{{}}-u_{j}^{{}}\right) ^{2}  \label{eq9.0}
\end{equation}

where $f_{lj}^{cc}=\partial ^{2}t_{lj}^{cc}/\partial \left(
u_{l}^{{}}-u_{j}^{{}}\right) ^{2}|_{0}$ $=f_{jl}^{cc}$ , etc. By defining $%
f_{k}^{cc}=z^{-1}\sum_{\langle j\rangle }f_{lj}^{cc}e^{ik\Delta _{lj}^{{}}}$
and the coefficients $F_{kpq}^{xy}=z\left(
f_{p}^{xy}+f_{k}^{xy}-2f_{k-q}^{xy}\right) $ with $x,y=c,d,$ the Fourier
transform reads: 
\[
\frac{1}{N}\sum_{kpq\sigma }c_{k\sigma }^{\dagger }c_{p\sigma
}^{{}}u_{q}^{{}}u_{k-p-q}^{{}}F_{kpq}^{cc}+\frac{1}{N}\sum_{kpq\sigma
}d_{k\sigma }^{\dagger }d_{p\sigma
}^{{}}u_{q}^{{}}u_{k-p-q}^{{}}F_{kpq}^{dd} 
\]
\begin{equation}
+\frac{1}{N}\sum_{kpq\sigma }\left( d_{k\sigma }^{\dagger }c_{p\sigma
}^{{}}u_{q}^{{}}u_{k-p-q}^{{}}+c_{p\sigma }^{\dagger }d_{k\sigma
}^{{}}u_{q}^{{}}u_{p-k-q}^{{}}\right) F_{kpq}^{cd}  \label{14.1}
\end{equation}

To pass over to the hybridized-fermion representation, let us define for
short: 
\begin{equation}
F_{kpq}^{\alpha \alpha }=F_{kpq}^{cc}\cos \varphi _{k}^{{}}\cos \varphi
_{p}^{{}}+F_{kpq}^{dd}\sin \varphi _{k}^{{}}\sin \varphi
_{p}^{{}}-F_{kpq}^{cd}\sin \varphi _{k}^{{}}\cos \varphi _{p}^{{}}
\label{18.1a}
\end{equation}

\begin{equation}
F_{kpq}^{\beta \beta }=F_{kpq}^{cc}\sin \varphi _{k}^{{}}\sin \varphi
_{p}^{{}}+F_{kpq}^{dd}\cos \varphi _{k}^{{}}\cos \varphi
_{p}^{{}}+F_{kpq}^{cd}\cos \varphi _{k}^{{}}\sin \varphi _{p}^{{}}
\label{18.1b}
\end{equation}

\begin{equation}
F_{kpq}^{\alpha \beta }=F_{kpq}^{cc}\cos \varphi _{k}^{{}}\sin \varphi
_{p}^{{}}-F_{kpq}^{dd}\sin \varphi _{k}^{{}}\cos \varphi
_{p}^{{}}-F_{kpq}^{cd}\sin \varphi _{k}^{{}}\sin \varphi _{p}^{{}}
\label{18.1c}
\end{equation}

\begin{equation}
F_{kpq}^{\beta \alpha }=F_{kpq}^{cc}\sin \varphi _{k}^{{}}\cos \varphi
_{p}^{{}}-F_{kpq}^{dd}\cos \varphi _{k}^{{}}\sin \varphi
_{p}^{{}}+F_{kpq}^{cd}\cos \varphi _{k}^{{}}\cos \varphi _{p}^{{}}
\label{18.1d}
\end{equation}

Then, the quadratic coupling can be cast in the form: 
\[
H_{ep}^{(2)}=\frac{1}{N}\sum_{kpq}\left( \alpha _{k\sigma }^{\dagger }\alpha
_{p\sigma }^{{}}F_{kpq}^{\alpha \alpha }+\beta _{k\sigma }^{\dagger }\beta
_{p\sigma }^{{}}F_{kpq}^{\beta \beta }+\alpha _{k\sigma }^{\dagger }\beta
_{p\sigma }^{{}}F_{kpq}^{\alpha \beta }+\beta _{k\sigma }^{\dagger }\alpha
_{p\sigma }^{{}}F_{kpq}^{\beta \alpha }\right) u_{q}^{{}}u_{k-p-q}^{{}} 
\]
\begin{equation}
+\frac{1}{N}\sum_{kpq}\left( \alpha _{p\sigma }^{\dagger }\alpha _{k\sigma
}^{{}}F_{kpq}^{\alpha \alpha }+\beta _{p\sigma }^{\dagger }\beta _{k\sigma
}^{{}}F_{kpq}^{\beta \beta }+\alpha _{p\sigma }^{\dagger }\beta _{k\sigma
}^{{}}F_{kpq}^{\beta \alpha }+\beta _{p\sigma }^{\dagger }\alpha _{k\sigma
}^{{}}F_{kpq}^{\alpha \beta }\right) u_{-q}^{{}}u_{-(k-p-q)}^{{}}
\label{20.1}
\end{equation}

Next, we quantize the phonons according to Eq.\ref{ph.quant}. By enforcing $%
k=p$ we take into account only the terms which do not change the number of
phonons in a given mode:\ 
\begin{equation}
u_{\pm q}^{{}}u_{\pm (k-p-q)}^{{}}\approx u_{q}^{{}}u_{-q}^{{}}\delta
_{pk}^{{}}=\delta _{pk}^{{}}L_{q}^{2}\left( b_{-q}^{\dagger }b_{q}^{\dagger
}+b_{q}^{{}}b_{-q}^{{}}+\nu _{q}^{{}}+\nu _{-q}^{{}}+1\right)  \label{20.3}
\end{equation}

Let us stress that our aim is to show that there are some contributions to $%
H_{ep}^{(2)}$ which provide an effective inter-band coupling. We do not
claim to be able to treat all the terms in $H_{ep}^{(2)}$: we just want to
select the subset of ''hot\ ''terms. Selecting the $p=q$ terms, then, Eq.\ref
{20.1} can be written compactly as 
\[
H_{ep}^{(2)}\approx \frac{1}{N}\sum_{kq}\left( 2F_{kkq}^{\alpha \alpha
}n_{k\sigma }^{\alpha }+2F_{kkq}^{\beta \beta }n_{k\sigma }^{\beta }\right)
L_{q}^{2}\left( b_{-q}^{\dagger }b_{q}^{\dagger }+b_{q}^{{}}b_{-q}^{{}}+\nu
_{q}^{{}}+\nu _{-q}^{{}}+1\right) 
\]
\begin{equation}
+\frac{1}{N}\sum_{kq}\left( F_{kkq}^{\beta \alpha }+F_{kkq}^{\alpha \beta
}\right) \left( \alpha _{k\sigma }^{\dagger }\beta _{k\sigma }^{{}}+\beta
_{k\sigma }^{\dagger }\alpha _{k\sigma }^{{}}\right) L_{q}^{2}\left(
b_{-q}^{\dagger }b_{q}^{\dagger }+b_{q}^{{}}b_{-q}^{{}}+\nu _{q}^{{}}+\nu
_{-q}^{{}}+1\right)  \label{21.2}
\end{equation}

\section{\protect\bigskip\ The electron-phonon Hamiltonian in the squeezed
phonon representation.}

Let us now introduce the squeezed phonon representation also for $%
H_{ep}^{(1)}+H_{ep}^{(2)}$. By using the relation $e^{S}\left(
b_{-q}^{\dagger }+b_{q}^{{}}\right) e^{-S}=e^{\eta _{q}^{{}}}\left(
b_{-q}^{\dagger }+b_{q}^{{}}\right) $ the linear coupling term becomes:

\[
e^{S}H_{ep}^{(1)}e^{-S}= 
\]
\begin{equation}
=\frac{1}{\sqrt{N}}\sum_{kq\sigma }L_{q}^{{}}e^{\eta _{q}^{{}}}\left[ \Gamma
_{k,k-q}^{\alpha \alpha }\alpha _{k\sigma }^{\dagger }\alpha _{k-q,\sigma
}^{{}}+\Gamma _{k,k-q}^{\beta \beta }\beta _{k\sigma }^{\dagger }\beta
_{k-q,\sigma }^{{}}+\Gamma _{k,k-q}^{\alpha \beta }\alpha _{k\sigma
}^{\dagger }\beta _{k-q,\sigma }^{{}}+\Gamma _{k,k-q}^{\beta \alpha }\beta
_{k\sigma }^{\dagger }\alpha _{k-q,\sigma }^{{}}\right] \left(
b_{-q}^{\dagger }+b_{q}^{{}}\right)  \label{SQ.ep1}
\end{equation}

Then the linear coupling has a reduce amplitude, as $\eta _{q}^{{}}<0$ (see
Eq.\ref{10.4}), consistently with the numerical analysis of Ref. \cite{choi}.

For the quadratic part $H_{ep}^{(2)}$ we get:

\[
e^{S}H_{ep}^{(2)}e^{-S}=\frac{1}{N}\sum_{kq}\left( 2F_{kkq}^{\alpha \alpha
}n_{k\sigma }^{\alpha }+2F_{kkq}^{\beta \beta }n_{k\sigma }^{\beta }\right)
L_{q}^{2}e^{2\eta _{q}^{{}}}\left( b_{-q}^{\dagger }b_{q}^{\dagger
}+b_{q}^{{}}b_{-q}^{{}}+\nu _{q}^{{}}+\nu _{-q}^{{}}+1\right) 
\]
\begin{equation}
+\frac{1}{N}\sum_{kq}\left( F_{kkq}^{\beta \alpha }+F_{kkq}^{\alpha \beta
}\right) \left( \alpha _{k\sigma }^{\dagger }\beta _{k\sigma }^{{}}+\beta
_{k\sigma }^{\dagger }\alpha _{k\sigma }^{{}}\right) L_{q}^{2}e^{2\eta
_{q}^{{}}}\left( b_{-q}^{\dagger }b_{q}^{\dagger }+b_{q}^{{}}b_{-q}^{{}}+\nu
_{q}^{{}}+\nu _{-q}^{{}}+1\right)  \label{23.1}
\end{equation}

The Eqs.\ref{SQ.ep1} \ and \ref{23.1} show that the phonons induce an
effective inter-band coupling through both the linear and the quadratic SSH
interaction. \ However, the quadratic term in the second line of Eq.\ref
{23.1} is the only one providing an \ inter-band coupling term which is
non-vanishing even if $\langle \nu _{q}^{{}}\rangle =0.$ It reads: 
\begin{equation}
L_{q}^{2}e^{2\eta _{q}^{0}}\left( F_{kkq}^{\beta \alpha }+F_{kkq}^{\alpha
\beta }\right) \left( \alpha _{k\sigma }^{\dagger }\beta _{k\sigma
}^{{}}+\beta _{k\sigma }^{\dagger }\alpha _{k\sigma }^{{}}\right)
\label{23.2}
\end{equation}

where $\tanh \left( 2\eta _{q}^{0}\right) =\lim_{\langle \nu
_{q}^{{}}\rangle \rightarrow 0}\tanh \left( 2\eta _{q}^{{}}\right) .$ It is
interesting to discuss its behaviour in the limit of small hybridization of
the bare bands. From the definition of $F_{kkq}^{\beta \alpha
},F_{kkq}^{\alpha \beta }$ (Eq.\ref{18.1c} and \ref{18.1d})\ and of $\varphi
_{k}^{{}}($ Eq.\ref{lin.7} ) one sees that, when the inter-band
hybridization $t_{lj}^{cd}\rightarrow 0,$ then either $\varphi
_{k}^{{}}\rightarrow 0,$ or $\varphi _{k}^{{}}\rightarrow \pm \pi /2.$ In
the two cases we have, from Eqs.\ref{18.1c} and \ref{18.1d}:

\begin{equation}
\lim_{\varphi _{k}^{{}}\rightarrow 0}F_{kkq}^{\alpha \beta }=0\qquad
\lim_{\varphi _{k}^{{}}\rightarrow 0}F_{kkq}^{\beta \alpha
}=F_{kkq}^{cd}\qquad \lim_{\varphi _{k}^{{}}\rightarrow \pm \pi
/2}F_{kkq}^{\alpha \beta }=-F_{kkq}^{cd}\qquad \lim_{\varphi
_{k}^{{}}\rightarrow \pm \pi /2}F_{kkq}^{\beta \alpha }=0  \label{23+1.1}
\end{equation}

The reason for the minus sign in $\lim_{\varphi _{k}^{{}}\rightarrow \pm \pi
/2}F_{kkq}^{\alpha \beta }=-F_{kkq}^{cd}$ can be understood by recalling
that, from Eq.\ref{eq.hybrid}, $\lim_{\varphi _{k}^{{}}\rightarrow \pm \pi
/2}c_{k\sigma }^{\dagger }\left( \varphi _{k}^{{}}\right) =\pm \beta
_{k\sigma }^{\dagger }$ while $\lim_{\varphi _{k}^{{}}\rightarrow \pm \pi
/2}d_{k\sigma }^{\dagger }\left( \varphi _{k}^{{}}\right) =\mp \alpha
_{k\sigma }^{\dagger }$ . It follows:

\begin{equation}
\lim_{\varphi _{k}^{{}}\rightarrow \pm \pi /2}\left( F_{kkq}^{\beta \alpha
}+F_{kkq}^{\alpha \beta }\right) \left( \alpha _{k\sigma }^{\dagger }\beta
_{k\sigma }^{{}}+\beta _{k\sigma }^{\dagger }\alpha _{k\sigma }^{{}}\right)
=-F_{kkq}^{cd}\left( -d_{k\sigma }^{\dagger }c_{k\sigma }^{{}}-c_{k\sigma
}^{\dagger }d_{k\sigma }^{{}}\right) =F_{kkq}^{cd}\left( d_{k\sigma
}^{\dagger }c_{k\sigma }^{{}}+c_{k\sigma }^{\dagger }d_{k\sigma }^{{}}\right)
\end{equation}

Hence in both cases $\varphi _{k}^{{}}\rightarrow 0,\pm \pi /2$ we get the
same expression for the effective inter-band coupling term, namely: 
\[
\lim_{t_{lj}^{cd}\rightarrow 0}\left( F_{kkq}^{\beta \alpha
}+F_{kkq}^{\alpha \beta }\right) \left( \alpha _{k\sigma }^{\dagger }\beta
_{k\sigma }^{{}}+\beta _{k\sigma }^{\dagger }\alpha _{k\sigma }^{{}}\right)
L_{q}^{2}e^{2\eta _{q}^{{}}}=F_{kkq}^{cd}L_{q}^{2}e^{2\eta _{q}^{{}}}\left(
d_{k\sigma }^{\dagger }c_{k\sigma }^{{}}+c_{k\sigma }^{\dagger }d_{k\sigma
}^{{}}\right) = 
\]
\begin{equation}
=L_{q}^{2}e^{2\eta _{q}^{{}}}\sum_{l\langle j\rangle }f_{lj}^{cd}\left\{
\cos \left( k\cdot \Delta _{lj}^{{}}\right) -\cos \left[ \left( k-q\right)
\cdot \Delta _{lj}^{{}}\right] \right\} \left( d_{k\sigma }^{\dagger
}c_{k\sigma }^{{}}+c_{k\sigma }^{\dagger }d_{k\sigma }^{{}}\right)
\end{equation}

Apart from geometric factors, this term depends only on the intensity of the
squeezing (through $e^{2\eta _{q}^{{}}})$ and on the amplitude of the
quadratic inter-band SSH electron-phonon coupling $f_{lj}^{cd}$. As $%
f_{lj}^{cd}$ is a second derivative of $t_{lj}^{cd},$ it can be appreciable
even if $t_{lj}^{cd}$ itself is very small. Different evaluations of $%
f_{lj}^{cd}$\cite{mazin} \ all agree that it has an appreciable value.

The weakening of the linear and quadratic electron-phonon interactions is
expressed by the coefficients $e^{\eta _{q}^{{}}}$ and $e^{2\eta _{q}^{{}}}$
. Their value is, in turn, set by the diagonalization condition of the
anharmonic phonon Hamiltonian, Eq.\ref{10.4} according to: 
\begin{equation}
e^{2\eta _{q}^{{}}}=\sqrt{\frac{1+\tanh \left( 2\eta _{q}^{{}}\right) }{%
1-\tanh \left( 2\eta _{q}^{{}}\right) }}
\end{equation}

Therefore the squeezing effect related to the anharmonicity of the phonons
also reduces the electron-phonon interactions.

To conclude, let us check if the link that our model establishes between the
softening of the harmonic frequency and the reduction of the electron-phonon
coupling strength is consistent with the estimates of those quantities as
given, e.g., in Refs.\cite{liu}, \cite{Yildirim} and \cite{choi} . The
softening of the frequency is evaluated \ as $15\%$.\cite{liu}, $25\%$\cite
{Yildirim} and $20\%$\cite{choi}. For our check we assume the intermediate
estimate by Ref.\cite{choi}. If $\omega _{q}^{{}}/X_{q}^{{}}\Omega
_{q}^{{}}=0.80$ then, from Eq.\ref{SQomega} , it follows ${\rm Th}\left(
2\eta _{q}^{{}}\right) =-0.60$ \ and therefore $e^{\eta _{q}^{{}}}=0.70$,
which agrees with the estimate \cite{choi} of a weakening of the linear
electron- phonon coupling by $30\%$.

\section{Conclusions.}

We have presented an analytic treatment of a two-band Hamiltonian with
anharmonic phonons and both linear and quadratic electron-phonon
interactions of the Su-Schrieffer-Heeger type, which should represent the
essential physics of $MgB_{2}.$ We have shown that the numerical results of
Ref.\cite{choi} about the phononic features of the material, namely the
softening of the effective harmonic frequency $\omega _{q}^{{}}$ and the
reduction of the linear electron-phonon coupling amplitude, can be
interpreted as due to the phonons accomodating themselves in a \ ''squeezed
''state. Additionally, we have found that the quadratic electron-phonon
interaction generates an effective coupling between the hybridized bands due
to virtual phonons.

\begin{acknowledgement}
We thank J. Spa\l ek and J. R. Iglesias for useful discussions.
\end{acknowledgement}

\end{document}